\documentclass[preprint]{revtex4}
\usepackage{graphicx}
\usepackage{epsfig}

\begin{document}

\title{Self-bound Interacting QCD Matter in Compact Stars}

\author{B. Franzon\dag\,  D. A. Foga\c{c}a\dag\,  F. S. Navarra\dag\ and J. E. Horvath\ddag\ }
\address{\dag\ Instituto de F\'{\i}sica, Universidade de S\~{a}o Paulo\\
Rua do Mat\~ao Travessa R, 187, 05508-090, S\~{a}o Paulo, SP, Brazil}
\address{\ddag\ Instituto de Astronomia, Geof\'isica e Ci\^encias Atmosf\'ericas -
Universidade de S\~ao Paulo, Rua do Mat\~ao, 1226, 05508-900, S\~ao Paulo, SP, Brazil}

\begin{abstract}

The quark gluon plasma (QGP) at zero temperature and high baryon number is a system that may be present
inside compact stars.  It is quite possible that this cold QGP shares some relevant features with the hot
QGP observed in heavy ion collisions, being also a strongly interacting  system. In a previous work we
have derived from the QCD Lagrangian an equation of state (EOS) for the cold QGP, which can be considered
an improved version of the MIT bag-model EOS. Compared to the latter, our EOS reaches higher
values of the pressure at comparable baryon densities. This feature is due to perturbative
corrections and also to nonperturbative effects. Here we apply this EOS to the study of neutron stars,
discussing the absolute stability of quark matter and computing the
mass-radius relation for self-bound (strange) stars. The maximum masses of the sequences exceed two solar masses,
in agreement with the recently measured
values of the mass of the  pulsar PSR J1614-2230, and the corresponding radii of around 10-11 km.

\end{abstract}

\maketitle



\vspace{1cm}
\section{Introduction}

In spite of the rapid progress in the field, the region of the QCD phase diagram with low temperature and high chemical
potential is still not well understood. According to the current status, supported by different model calculations, there is
a low temperature deconfined phase of quarks and gluons, the cold QGP, in which we may have color superconducting phases \cite{color}.
One of the open questions concerning the cold QGP is how free are quarks and gluons in this phase. In the simple picture
based on the MIT bag-model quarks and gluons do not interact except when they hit the bag wall. In the opposite corner of the
QCD phase diagram, i.e., at zero chemical potential and high temperature, the equivalent picture of a hot ideal gas of
noninteracting quarks and gluons was dramatically changed after the series of experiments with relativistic heavy-ion collisions
conducted at RHIC and now at LHC. In the new picture, quarks and gluons form a strongly interacting system in which nonperturbative
physics persist even after twice the critical temperature. In particular, the gluon condensates do not disappear very rapidly as previously expected.
In the case of the cold QGP, experiments in laboratories cannot be performed directly, since compression of
cold nuclear matter up to these tremendous densities cannot be achieved.
However this compression occurs presumably in the core of dense stars and the idea
that we might find cold QGP in neutron stars has been around already for some decades \cite{decades,alford,li}.
It is even conceivable that a whole star,
not only its core, be made of quark matter \cite{witten84}. This possibility will be explored in this work.

The existence of a deconfined quark phase in the core of neutron stars \cite{alford,li} depends crucially on the EOS.
On the theoretical side there is still considerable freedom, since it is easy to calculate the mass and the radius of a star for a given
EOS. Changing parameters in the proposed EOS one can arrive at  rather different curves in the mass-radius diagram. On the
observational side it is very difficult to obtain the mass and the radius of one single object. However, once this combined information is available
it will provide a very strong constraint on the EOS of dense matter.
The most recent data already impose some limits on the EOS parameters
\cite{pagliara2011,horvath2011,sergioduarte2010}.
A previous analysis of the observational data from the neutron star EXO 0748-676 presented in \cite{ozel2006} concluded that  most of the
EOS are too soft and therefore unable to support the existence of  neutron stars  with a quark phase. In spite of this
conclusion being disputed \cite{pagliara2011}, new precise measurements of the pulsar PSR J1614-2230 carried out in \cite{nature} yielded
a mass of $1.97 \pm 0.03 M_{\odot}$ for this object and led to the idea of interacting quarks if a core is present \cite{ozeldemorest}. Alternatively, a self-bound star, composed entirely of quark matter, could
explain a massive pulsar if the pairing interactions and vacuum energy fall in the right range \cite{horvath2011}, depending
on the value of the radius which is still under discussion.
It is then interesting to explore the existence of a self-bound deconfined quark phase made of noninteracting quarks \cite{nature}, as
suggested in previous attempts \cite {pagliara2011, prakash2010}.

In this paper we consider a quark star consisting of {\it u}, {\it d} and {\it s} quarks.
Heavier quarks are not present in neutron stars
\cite{fridolinlivro}. We shall further assume that the masses of the quarks
are $m_{u}=$ $5$ $MeV$, $m_{d}=$ $7$ $MeV$, and $m_{s}=$ $150$ $MeV$, complying with the generally accepted assumption
of two light flavors and a heavier $s$ quark.
We first study the absolute stability parameter space of the EOS derived in \cite{davi}, which describes the  quark gluon plasma at zero temperature.
Stability requirements restrict the range of parameter values, which are subsequently used in the construction of the mass-radius diagram.

This text is organized as follows. In Sec. II we briefly review the EOS for the cold QGP. In Sec. III we introduce the
stability conditions and discuss its consequences. In Sec. IV we present the Tolman-Oppenheimer-Volkoff (TOV) equations for stellar
structure calculations and their numerical solutions.  Finally in Sec. V we present some comments and conclusions.

\section{EOS of the cold QGP }

In \cite{davi} EOS derivation started with the assumption that the gluon field can be decomposed into low (``soft'') and high
(``hard'') momentum components. The expectation values of the soft fields were identified with the gluon condensates of dimension two and
four, respectively. The former generates a dynamical mass, $m_G$ for the hard gluons, and the latter yields an analogue of the ``bag constant'' term
in the energy density and pressure.  Given the large number of quark sources, even in the weak coupling regime, the hard gluon fields
are strong, the occupation numbers are large, and therefore these fields can be approximated by classical color fields.
The effect of the condensates is to soften the EOS whereas the hard gluons significantly stiffens it, by increasing both the energy density and pressure. With these approximations it was possible to derive \cite{davi} an analytical expression for the EOS, called here MFTQCD
(Mean Field Theory of QCD).

To proceed for the stellar conditions, we consider quarks $u$, $d$, $s$ and electrons in chemical equilibrium maintained by the weak processes \cite{farhi}:
$$
u+e^{-} \rightarrow d + \nu_{e},\hspace{1.5cm}
u+e^{-} \rightarrow s + \nu_{e},
$$
\begin{equation}
d \rightarrow u+e^{-} + \bar{\nu}_{e},\hspace{1.5cm}
s \rightarrow u+e^{-} + \bar{\nu}_{e},  \hspace{1cm} \textrm{and} \hspace{1cm}
s + u\rightarrow d+ u.
\label{w}
\end{equation}
As usual, the neutrinos are assumed to escape and do not contribute to the pressure and energy density.
In chemical equilibrium we have
\begin{equation}
\mu_{d} = \mu_{s} \equiv \mu
\hspace{1cm} \textrm{and} \hspace{1cm}
\mu_{d} + \mu_{e}= \mu.
\label{ce}
\end{equation}
The charge neutrality and baryon number conservation require
\begin{equation}
{\frac{2}{3}}\rho_{u}={\frac{1}{3}}\rho_{d}+{\frac{1}{3}}\rho_{s}+\rho_{e},
\label{cn}
\end{equation}
and
\begin{equation}
\rho_{B}={\frac{1}{3}}(\rho_{u}+\rho_{d}+\rho_{s}),
\label{bc}
\end{equation}
where $\rho_{B}$ is the total baryon density and $\rho_i$ is the density of quarks of flavor $i$ $({i=u,d,s})$
defined by the corresponding Fermi momentum $k_{i}$ given by
\begin{equation}
\rho_{i}=\frac{\gamma_{Q}}{2\pi^2}k_{i}^3
\label{densidade}
\end{equation}
(note that we impose a local conservation of the charges). The electron density is
\begin{equation}
\rho_{e}=\frac{\gamma_{e}}{6\pi^2}k_{e}^3,
\label{densidadee}
\end{equation}
where $\gamma_{Q}$ and $\gamma_{e}$ are the quark and electron degeneracy factors given by $\gamma_{Q} = \gamma_{e} = 2$
due to spin (the sum over color states was already performed). From (\ref{w}) to (\ref{densidadee}) we find a
set of four algebraic equations for Fermi momentum calculation for each particle:
$$
{k_{u}}^{3}+{k_{d}}^{3}+{k_{s}}^{3}=3\pi^{2}\rho_{B},
$$
\begin{equation}
2{k_{u}}^{3}={k_{d}}^{3}+{k_{s}}^{3}+{k_{e}}^{3},
\label{densidadee}
\end{equation}
$$
{k_{d}}^{2}+{m_{d}}^{2}={k_{s}}^{2}+{m_{s}}^{2},
$$
$$
\sqrt{{k_{u}}^{2}+{m_{u}}^{2}}+\sqrt{{k_{e}}^{2}+{m_{e}}^{2}}=\sqrt{{k_{s}}^{2}+{m_{s}}^{2}},
$$
for a fixed baryon density $\rho_{B}$. The energy density is finally given by \cite{davi}
$$
\varepsilon=\bigg({\frac{27g^{2}}{16{m_{G}}^{2}}}\bigg) \ {\rho_{B}}^{2} +   \mathcal{B}_{QCD}
$$
$$
+\sum_{i=u,d,s}3{\frac{\gamma_{Q}}{2{\pi}^{2}}} \Bigg\lbrace {\frac{{k_{i}}^{3}\sqrt{{k_{i}}^{2}+{m_{i}}^{2}}}{4}} +
{\frac{{m_{i}}^{2}{k_{i}}\sqrt{{k_{i}}^{2}+{m_{i}}^{2}}}{8}} - {\frac{{m_{i}}^{4}}{8}}ln\Big[{k_{i}}+\sqrt{{k_{i}}^{2}+{m_{i}}^{2}} \ \Big] + {\frac{{m_{i}}^{4}}{16}}ln({m_{i}}^{2}) \Bigg\rbrace
$$
\begin{equation}
+{\frac{\gamma_{e}}{2{\pi}^{2}}} \Bigg\lbrace {\frac{{k_{e}}^{3}\sqrt{{k_{e}}^{2}+{m_{e}}^{2}}}{4}} +
{\frac{{m_{e}}^{2}{k_{e}}\sqrt{{k_{e}}^{2}+{m_{e}}^{2}}}{8}} - {\frac{{m_{e}}^{4}}{8}}ln\Big[{k_{i}}+\sqrt{{k_{e}}^{2}+{m_{e}}^{2}} \ \Big] + {\frac{{m_{e}}^{4}}{16}}ln({m_{e}}^{2}) \Bigg\rbrace,
\label{epsib}
\end{equation}
and the pressure is
$$
p=\bigg({\frac{27g^{2}}{16{m_{G}}^{2}}}\bigg) \ {\rho_{B}}^{2}  -   \mathcal{B}_{QCD}
$$
$$
+\sum_{i=u,d,s}{\frac{\gamma_{Q}}{2{\pi}^{2}}} \Bigg\lbrace {\frac{{k_{i}}^{3}\sqrt{{k_{i}}^{2}+{m_{i}}^{2}}}{4}}  -
{\frac{3{m_{i}}^{2}{k_{i}}\sqrt{{k_{i}}^{2}+{m_{i}}^{2}}}{8}} + {\frac{3{m_{i}}^{4}}{8}}ln\Big[
{k_{i}}+\sqrt{{k_{i}}^{2}+{m_{i}}^{2}} \ \Big]-{\frac{3{m_{i}}^{4}}{16}}ln({m_{i}}^{2}) \Bigg\rbrace
$$
\begin{equation}
+{\frac{\gamma_{e}}{6{\pi}^{2}}} \Bigg\lbrace {\frac{{k_{e}}^{3}\sqrt{{k_{e}}^{2}+{m_{e}}^{2}}}{4}} -{\frac{3{m_{e}}^{2}{k_{e}}\sqrt{{k_{e}}^{2}+{m_{e}}^{2}}}{8}} + {\frac{3{m_{e}}^{4}}{8}}ln\Big[
{k_{e}}+\sqrt{{k_{e}}^{2}+{m_{e}}^{2}} \ \Big]-{\frac{3{m_{e}}^{4}}{16}}ln({m_{e}}^{2}) \Bigg\rbrace ,
\label{pressb}
\end{equation}
where $m_{e}=$ $0.5$ $MeV$ is the electron mass, $m_{G}$ is the dynamical gluon mass, and $g$ is the coupling constant $(\alpha_{s}=g^{2}/4\pi)$ in QCD.
Our analogue of the bag constant, called here $\mathcal{B}_{QCD}$, is given by
\begin{equation}
\mathcal{B}_{QCD}= \frac{9}{128} \, \phi_{0}^4 = \langle \frac{1}{4} F^{a\mu\nu}F^{a}_{\mu\nu} \rangle,
\label{bag}
\end{equation}
where $\phi_0$ is an energy scale associated with the energy density of the vacuum and with the gluon condensate \cite{davi}.
In (\ref{epsib}) and (\ref{pressb}) the summation over quark colors has already been performed.
Throughout this work we employ the natural units $\hbar=1$, $c=1$.

\section{Stability conditions for the EOS}

We are interested in studying star models with stable strange quark matter.  In this case, we have two stability conditions.
The first one is that  the energy per baryon of the deconfined phase (for $P =0$ and $T=0$) is lower than the nonstrange infinite baryonic matter defined in \cite{farhi,pagliara2011}. Following these works we impose that:
\begin{equation}
E_{A} \equiv \frac{\varepsilon}{\rho_{B}} \leq  934  \,\,\,  \mbox{MeV}.
\label{estabilidade}
\end{equation}
Since this condition must hold at the zero pressure point, from  (\ref{epsib}) and (\ref{pressb}) we can numerically
derive a relation between the bag constant $B_{QCD}$ and the ratio $\xi=g/m_{G}$. We solve  (\ref{pressb})  obtaining
$\rho_B=\rho_B (B_{QCD}, \xi)$, which is then inserted into  (\ref{epsib}). The resulting expression is used to write the
condition $\varepsilon (B_{QCD}, \xi)/ \rho_B (B_{QCD}, \xi) = 934$ $MeV$, which defines one `` stability frontier ''.
This last equation is rewritten as $\xi = \xi (B_{QCD})$, is plotted in Fig. 1 (solid line) and denoted by the 3-flavor line.
Points in the
$ \mathcal{B}_{QCD} \, - \, \xi$ plane located on the right of the solid line are discarded since they do not satisfy
(\ref{estabilidade}). The solid line, corresponding to the maximal value of $E_{A} = 934$ $MeV$, determines the maximum value of
$\mathcal{B}_{QCD} \simeq 75.7$   $MeV /fm^{3}$. The minimum value of $\mathcal{B}_{QCD} \simeq 38$  $MeV/fm^3$ is determined by the
second stability condition, which requires nonstrange
quark matter in the bulk to have an energy per baryon higher than the one of nonstrange infinite baryonic matter. By imposing that
\begin{equation}
E_{A} \equiv \frac{\varepsilon}{\rho_{B}}   \geq  934  \,\,\,  \mbox{MeV}
\label{estabilidade2}
\end{equation}
for  a two flavor quark matter at ground state, we ensure that atomic nuclei do not dissolve into their constituent
quarks. The constraint (\ref{estabilidade2})  defines the dotted line in the $ \mathcal{B}_{QCD} \, - \, \xi$ plane, denoted by the
2-flavor line in Fig. 1.  Points located on the left of this line are excluded because they do not satisfy (\ref{estabilidade2}).
The region between the two lines in Fig. 1 defines our stability window.

\begin{figure}[h]
\begin{center}
\epsfig{file=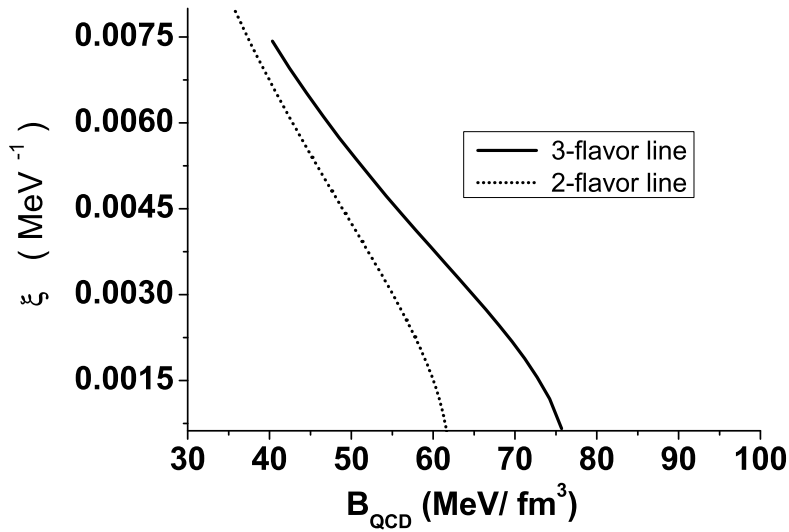,width=116mm}
\caption{Values of  $\xi = g/m_{G}$ as a function of  $\mathcal{B}_{QCD}$ for different values of the energy per baryon. }
\end{center}
\label{fig1}
\end{figure}
The requirement of strange quark matter stability at finite pressure, in the interior of the stars, demands the introduction
of another criterion. We shall assume that among the quark matter phase and the hadron phase, represented here by two hadronic
models, the most stable is the one which has the  highest pressure for the same value of the chemical potential. The curves $p$ versus $\mu_B$ obtained with  three EOS are shown in Fig. 2. As can be seen, our quark matter
is more stable than the matter described by the hadronic models studied here.

\begin{figure}[h]
\begin{center}
\epsfig{file=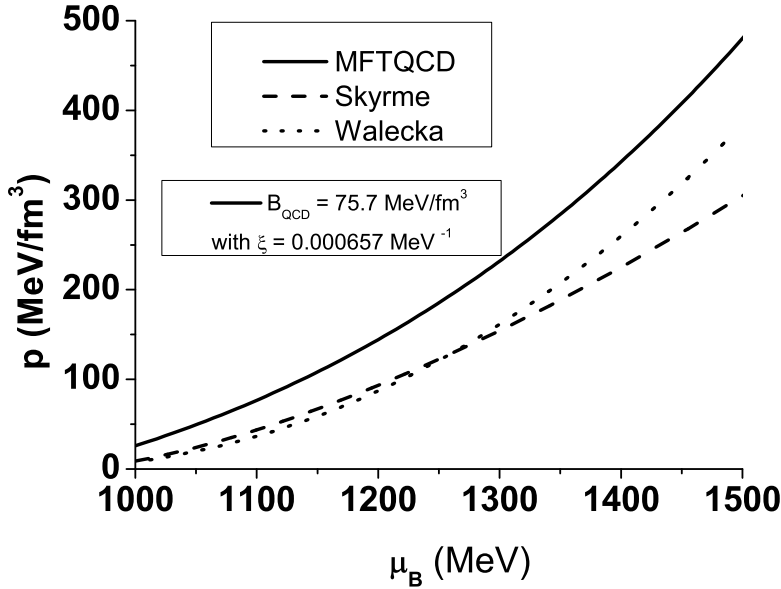,width=116mm}
\caption{Pressure as a function of the chemical potential for the three EOS:
MFTQCD \cite{davi} ,
Skyrme \cite{douchin}, and Walecka \cite{glend}.}
\end{center}
\label{fig2}
\end{figure}

From Fig. 2 we can conclude that, at increasing chemical potential (and density),  quark matter becomes more and
more favored  with respect to the hadronic matter studied here.

We performed the causality check for $\mathcal{B_{QCD}}=38$ $MeV/fm^{-3}$ (close to the minimum value) and for (the maximum value)  $\mathcal{B_{QCD}}=75.7$ $MeV/fm^{-3}$. These two values of the bag constant define the stability range.
Using  these two values in Fig. 1 as entries to the dotted line and to the solid line, respectively, we can read in the vertical axis the two
corresponding values of the variable $\xi$, which are   $\xi=0.007293$ $MeV^{-1}$
for $\mathcal{B_{QCD}}=38$ $MeV/fm^{-3}$ and $\xi=0.000657$ $MeV^{-1}$ for $\mathcal{B_{QCD}}=75.7$ $MeV/fm^{-3}$.
Having fixed these parameters, we go back to (\ref{epsib}) and (\ref{pressb}) and, obtaining $\varepsilon$ and $p$ for successive values of $\rho_B$,
we construct the EOS in the  form $p = p(\varepsilon)$, plotted in Fig. 3. In this type of plot the slope is the speed of sound,
which, due to causality, can not exceed the unity. This limit is shown by the full lines in the figure.

\begin{figure}[h]
\begin{center}
\epsfig{file=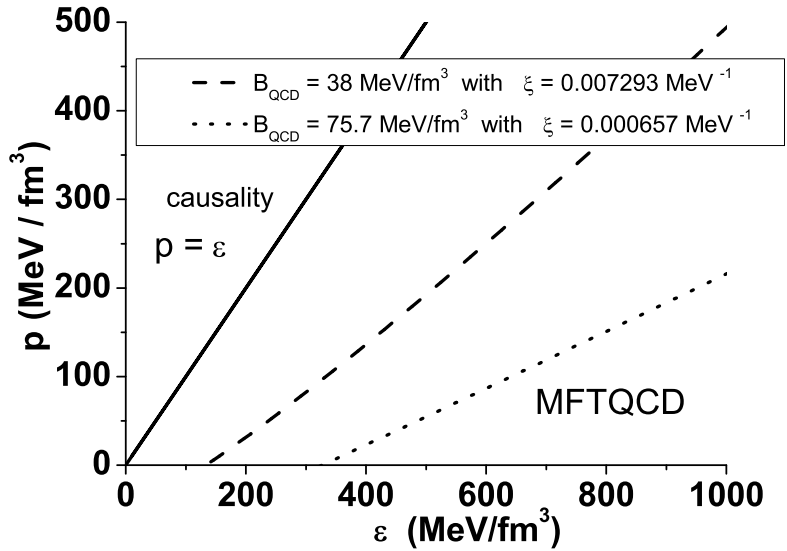,width=116mm}
\caption{EOS for the cold quark--gluon plasma.}
\end{center}
\label{fig3}
\end{figure}

\section{Numerical solutions of the  TOV equation}

In order to describe the structure of a static, non-rotating compact star we solve the Einstein equations \cite{OPP}:
\begin{equation}
G^{\mu\nu}=-8\pi GT^{\mu\nu},
\label{ee}
\end{equation}
for a spherical, isotropic, static, and  general relativistic ideal fluids in hydrostatic equilibrium. This particular solution of (\ref{ee}) leads to
the Tolman-Oppenheimer-Volkoff (TOV) equation for the pressure $p(r)$:
\begin{equation}
\frac{dp}{dr}=-\frac{G \epsilon(r) M (r)}{r^2} \left[ 1 + \frac{p(r)}{\epsilon(r)} \right] \left[ 1 + \frac{4\pi r^3 p(r)}{M(r)} \right] \times
\left[ 1 - \frac{2GM(r)}{r} \right]^{-1},
\label{tov}
\end{equation}
where $G$ is the Newton gravitational constant. The enclosed mass $M(r)$ of the compact star is given by the mass continuity equation:
\begin{equation}
\frac{dM(r)}{dr}=4\pi r^2\epsilon(r).
\label{mass}
\end{equation}
Equations (\ref{tov}) and (\ref{mass}) express the balance between the gravitational force and the internal pressure acting on a shell of mass
$dM(r)$ and thickness $dr$.

We solve numerically (\ref{tov}) and (\ref{mass}), which are coupled nonlinear equations for $p(r)$ and $M(r)$, to obtain the mass-radius diagram.
The pressure and the energy density in (\ref{tov}) and (\ref{mass}) are given by the MFTQCD expressions (\ref{pressb}) and (\ref{epsib}), respectively.
We take the central energy density to be  $\epsilon(r = 0)=\epsilon_{c}$ and then we integrate out (\ref{tov}) and (\ref{mass}) from $r=0$ up to $r=R$, where the pressure at the surface is zero: $p(r=R)=0$. In Fig. 4 we show the mass-radius diagram for several values of  $\mathcal{B_{QCD}}$ and $\xi$ respecting the stability condition.

\begin{figure}[h]
\begin{center}
\epsfig{file=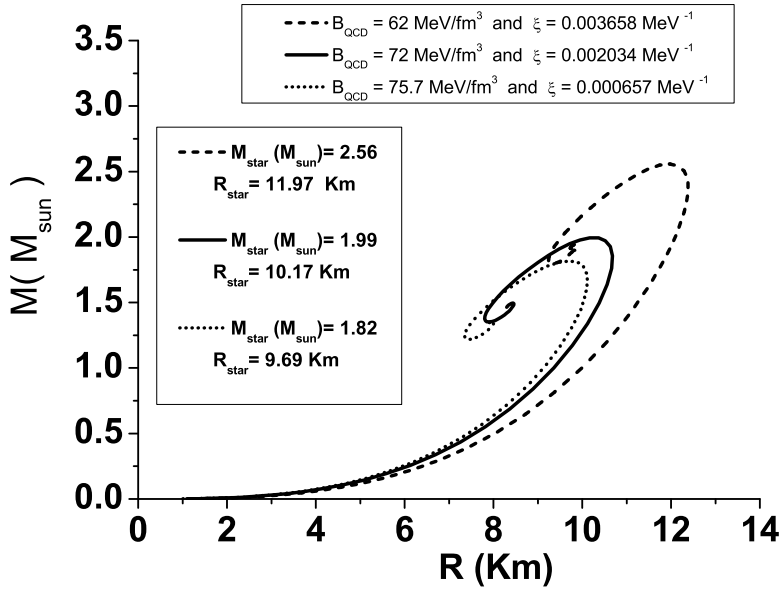,width=116mm}
\caption{Mass-radius diagram for several values of $B_{QCD}$ and $\xi$ allowed by the stability condition.}
\end{center}
\label{fig4}
\end{figure}
Table \ref{quarstar} summarizes the values of mass, radius, and central energy density obtained for the several values of  $\mathcal{B_{QCD}}$
shown in  Fig. 4. At this point, the relationship of this parameter to other commonly employed quantities (i.e. the MIT bag constant) is
difficult to assess, and the reasonable values obtained for the stability window to hold are quite encouraging.

\begin{table}[!htbp]
\caption{Bag, Maximum Mass, and Radius of the quark star.}
\vspace{0.3cm}
\centering
\begin{tabular}{cccc}
\hline
$\mathcal{B_{QCD}} (MeV/fm^{3})$ & $\xi (MeV^{-1}) $  & M$ (M_{\odot})$ & R$(Km)$ \\ [0.8ex]
\hline
\hline
62 &  0.003658  &   2.56 & 11.97 \\
\hline
72 &  0.002034  &  1.99  & 10.17  \\
\hline
75.7 &  0.000657   & 1.82   & 9.69  \\
\hline
\end{tabular}
\label{quarstar}
\end{table}

\section{Conclusion}

In this paper we have applied an EOS of the cold QGP to the study of compact stars. We note that when gluon interactions
are switched off, we recover the standard MIT bag model EOS. The inclusion of gluon interactions generates more pressure and
energy density, rendering the equation
of state harder than the MIT bag model one and able to support stellar sequences with larger maximum masses.
Indeed, our solutions of the TOV equations yield
stars with two solar masses, in agreement with recent observations \cite{nature}. In the present paper we have improved a previous one
\cite{eua} in several aspects. The most important one was to introduce the requirement of stability, which strongly constrained the range
of possible parameters. However, even after this strong restriction of parameter choice, we were still able to find stable quark stars with
acceptable masses and radii. The latter is never too large ($R \leq 12 km$), even for stellar sequences with maximum masses of $\sim 2.5 M_{\odot}$,
therefore determinations of radii underway \cite{ozelguver,guvercamarota,steinerlatt} have the potential of constraining or even ruling out this type of theory in the near future.

\begin{acknowledgments}
We are deeply grateful to Richard R. Silbar and Sergio B. Duarte for fruitful discussions.
This work was partially financed by the Brazilian funding agencies CAPES, CNPq and FAPESP.
\end{acknowledgments}

\end{document}